%% file: conference_101719.tex
\documentclass[conference]{IEEEtran}
\IEEEoverridecommandlockouts
\usepackage{cite}
\usepackage{amsmath,amssymb,amsfonts}
\usepackage{algorithmic}
\usepackage{graphicx}
\usepackage{textcomp}
\usepackage{xcolor}
\usepackage{multirow}
\def\BibTeX{{\rm B\kern-.05em{\sc i\kern-.025em b}\kern-.08em
    T\kern-.1667em\lower.7ex\hbox{E}\kern-.125emX}}
\begin{document}
\title{AutoStyle-TTS: Retrieval-Augmented Generation based Automatic Style Matching Text-to-Speech Synthesis}
% Retrieval-Augmented Generation based Automatic Style Matching Text-to-Speech synthesis

% \author{\IEEEauthorblockN{1\textsuperscript{st} Dan Luo$^{\ast}$\thanks{$^{\ast}$  Equal Contribution}$^{\ddagger}$ \thanks{$^{\ddagger}$ Work done during internship at Tencent AI Lab.}}
% \IEEEauthorblockA{
% \textit{Tsinghua University}\\
% Shenzhen, China \\
% luod23@mails.tsinghua.edu.cn}
% \and
% \IEEEauthorblockN{2\textsuperscript{nd} Chengyuan Ma$^{\ast}$$^{\ddagger}$}
% \IEEEauthorblockA{
% \textit{Tsinghua University}\\
% Shenzhen, China \\
% mcy23@mails.tsinghua.edu.cn}
% \and
% \IEEEauthorblockN{3\textsuperscript{rd} Weiqin Li}
% \IEEEauthorblockA{
% \textit{Tsinghua University}\\
% Shenzhen, China \\
% lwq22@mails.tsinghua.edu.cn}
% \and
% \IEEEauthorblockN{4\textsuperscript{th} Jun Wang$^{\dagger}$\thanks{$^{\dagger}$ Corresponding Author}}
% \IEEEauthorblockA{
% \textit{Tencent AI Lab} \\
% % \textit{}\\
% Shenzhen, China \\
% joinerwang@tencent.com}
% \and
% \IEEEauthorblockN{5\textsuperscript{th} Wei Chen}
% \IEEEauthorblockA{
% \textit{Tsinghua University}\\
% Shenzhen, China \\
% chenw23@mails.tsinghua.edu.cn}
% \and
% \IEEEauthorblockN{6\textsuperscript{th} Zhiyong Wu$^{\dagger}$}
% \IEEEauthorblockA{
% \textit{Tsinghua University}\\
% Shenzhen, China \\
% zywu@sz.tsinghua.edu.cn}
% }

\author{
\IEEEauthorblockN{
  Dan Luo $^{1,\ast}$,
  $^{1}$Chengyuan Ma$^{1,\ast}$,
  Weiqin Li$^{1}$,
  Jun Wang$^{2,\dagger}$,
  Wei Chen$^{1}$,
  Zhiyong Wu$^{1,\dagger}$
}
\IEEEauthorblockA{
  \textsuperscript{1}\textit{Shenzhen International Graduate School, Tsinghua University, Shenzhen, China}\\
  \textsuperscript{2}\textit{AI Lab, Tencent, Shenzhen, China}\\
  \texttt{\{luod23,mcy23\}@mails.tsinghua.edu.cn}
}
\thanks{$^{\ast}$ Equal Contribution}
% \thanks{$^{\ddagger}$ Work done during internship at Tencent AI Lab.}
\thanks{$^{\dagger}$ Corresponding Author}
}

\maketitle

\begin{abstract}
With the advancement of speech synthesis technology, users have higher expectations for the naturalness and expressiveness of synthesized speech. But previous research ignores the importance of prompt selection. This study proposes a text-to-speech (TTS) framework based on Retrieval-Augmented Generation (RAG) technology, which can dynamically adjust the speech style according to the text content to achieve more natural and vivid communication effects. We have constructed a speech style knowledge database containing high-quality speech samples in various contexts and developed a style matching scheme. This scheme uses embeddings, extracted by Llama, PER-LLM-Embedder,and Moka, to match with samples in the knowledge database, selecting the most appropriate speech style for synthesis. Furthermore, our empirical research validates the effectiveness of the proposed method. Our demo can be viewed at: https://thuhcsi.github.io/icme2025-AutoStyle-TTS
\end{abstract}


\section{Introduction}
\label{sec:intro}
\input{sections/intro}
\section{METHODOLOGY}
\input{sections/method}

\section{EXPERIMENTS AND RESULTS}
\input{sections/exp}

\section{conclusion}
In this paper, we have developed a novel TTS framework that integrates RAG technology. This framework is designed to dynamically adjust speech styles in response to textual prompts and user preference, resulting in a more natural and engaging auditory experience. We construct a comprehensive speech style database, combined with our style matching scheme and embedding extractor, allowing for the precise selection and application of speech styles that align with the contextual requirements of the input text. Our empirical evaluations have demonstrated that our get excellent effect in terms of the style matching between text and speech and the style coherence in speech.

\section{Acknowledgment}
This work is supported by National Natural Science Foundation of China (62076144) and Shenzhen Science and Technology Program (JCYJ20220818101014030).

\bibliographystyle{IEEEbib}
\bibliography{refs,refs2}

\vspace{12pt}
\color{red}

\end{document}

%% file: sections/intro.tex
%para_1
Existing language model (LM) based TTS models \cite{wang2023neuralcodeclanguagemodels,xie2024controllablespeechsynthesisera,du2024cosyvoice,yang2024simplespeech2simpleefficient,yang2023instructttsmodellingexpressivetts} have achieved highly realistic results in speech generation and voice cloning, and podcast synthesis is one of the most important application scenarios\cite{10096247}.
%lwq1221 Emotional fluctuations是不是好一点 （✅）
Whether it is storytelling or interview dialogue, an engaging podcast is often accompanied by changes of subject, interactions between speakers, and emotional fluctuations\cite{bennett2013logic}.
%lwq1221 这里的选择 是否换成 调整 好一些 （✅）
As a result, speech synthesis technology must not only be able to control the speaker's timbre and style of speech, but also be able to dynamically adapt the style of speech according to different contextual situations in order to create more natural and immersive user experience.

% para2
%lwq1221 这里感觉缺少一种递进的关系。或许可以改成：近来一些工作通过加入文本instrcut和语音prompt的方式让TTS合成特定风格的语音，从而应用于需要风格变化的场景。cosyvoice是用文本tag来建模特定现象，但是缺少对特定现象外的全局语音风格的控制（我觉得第一个例子你可以换一下，参考ustyle，他说的好像是用比如emotion tag这种tag来代表风格？我不确定这种特定现象的tag放在这边是否合适）；voxinstruct使用文本prompt来描述风格，用文本描述音色、情感（emotion）等信息来控制整个语音的风格，但是文本很难准确地描述语音风格；stylefusion使用语音prompt提取的语音风格信息作为指导，实现了较好的风格语音合成。
% 但是这些方法都只关注于对合成语音风格的控制，忽略了错误的风格提示对合成语音表现力的影响。尤其是在博客场景下，风格的丰富变化使得对风格提示的准确性/正确性 有更高的要求，同时自动选择合适的风格也能降低手动选择带来的额外成本。（分别对应 Content and style disharmony 和高校选择风格）
Several studies have begun to explore how to achieve accurate and fine control over speech style, which can be implemented through text instructions or by modeling speech styles from speech prompts. Voxinstruct~\cite{zhou2024voxinstruct} uses text to describe the style, specifying information such as timbre and mood as textual prompts for overall speech style control; CosyVoice~\cite{du2024cosyvoice}, U-style~\cite{li2024u} and StyleFusion~\cite{chen2024stylefusion} extract speech style representations from speech prompts with specific style extractors, such as redesigned U-net.
One key feature of these methods is that the generation quality of these GPT-like generative models depends heavily on the selection and design of the prompt~\cite{wang2024emopropromptselectionstrategy}.
% 然而，这些方法都只关注于对合成语音风格的控制，忽略了错误的风格提示对合成语音表现力的影响。尤其是在博客场景下，风格的丰富变化使得对风格提示的准确性/正确性 有更高的要求，同时自动选择合适的风格也能降低手动选择带来的额外成本。（分别对应 Content and style disharmony 和高校选择风格）
However, all of these approaches only focus on controlling the style of the synthesized speech, ignoring the effect of incorrect stylistic cues on the expressiveness of the synthesized speech.~\cite{bott2024controllingemotiontexttospeechnatural}. Especially in podcasting scenarios, the rich variation of styles makes the correctness of style cues more demanding, and the automatic selection of appropriate styles can also reduce the extra cost of manual selection.

% An appropriate prompt can more accurately guide the speech synthesis system to capture and reproduce the target speech style, while an inappropriate prompt may lead to a deviation of the speech output from the expected style, affecting the final user experience~\cite{bott2024controllingemotiontexttospeechnatural}. 

%lwq1221 这里One key这里 和 previous work suffers这里有重复。都是分析之前的不足，可以放在一段。并且这里 tend to model only one fixed style 是否准确？前面的工作好像不是fixed style，咱这个工作的点应该在于自动选择风格 而不是手动选择。是不是inefficient更合适？
More specifically, previous work suffers from the following problems in prompt selection: \textbf{(i) Limited availability.} Conventional TTS~\cite{zhou2024voxinstruct} systems tend to model the style provided by text or speech and then apply this style to all generated speech. These approaches expose obvious limitations  The traditional approach makes it difficult to realise the flexibility of adjusting the tone, pace and emotional expression when facing long text-generated and dialogue-generated podcast scenarios.
% These approaches exposes obvious drawbacks when faced with long texts or dialogue scenarios. For example, when narrating a story, the narrator needs to flexibly adjust the tone, tempo, and emotional expression according to different plot developments, which is difficult for traditional approaches to achieve. 
In addition, manually selecting the appropriate voice style for each sentence is not only inefficient but also leads to inconsistencies between the content and the actual style of expression. \textbf{(ii) Content and style disharmony.} The modeled speech styles of these LLM-based speech synthesis approaches~\cite{du2024cosyvoice,li2024u,chen2024stylefusion} do not effectively take into account the specific content of the text as well as the relevance of the contextual scenarios. This leads to a lack of natural variation and coherence between synthesized speech.

%para3
%lwq1223 rag的优点在于让模型能处理复杂的查询从而生成准确的信息？这里能否扯一下rag的优点是 在特定的场景下通过检索生成更好的结果，突出我们是为了 符合上下文场景 而使用的rag。 最后一句用 which has been proven effective in the domains of text and image generation.
In the field of prompt enhancement, Retrieval-Augmented Generation(RAG) significantly improves the model's ability to handle complex queries and generate more accurate information by introducing an external knowledge database and comprehensively analyzing specific scenario information~\cite{gao2023retrieval,lewis2021retrievalaugmentedgenerationknowledgeintensivenlp}, which has been proven effective in the domains of text, image and audio generation~\cite{christmann2024ragbasedquestionansweringheterogeneous,yuan2024retrievalaugmentedtexttoaudiogeneration,choi24c_interspeech,xue2024retrieval}. Inspired by this, and in order to solve the problem of limited availability, we introduce the RAG technique to allow the model to automatically select the appropriate speech style prompt to guide the style of the synthesized speech without the need for manual selection. The application of RAG also greatly enhances the flexibility of the speech styles so that the synthesised speech can be flexibly varied according to different scenarios and needs.

In this paper, we propose a novel TTS framework, AutoStyle-TTS, which addresses the previous system's shortcomings in usability and the disconnect between content and style. Considering the ability to automatically select matching style prompts, we introduced RAG. Due to the limitations of existing speech style knowledge databases and the need to consider the correlation between text style and content, we designed a method for constructing a speech style knowledge database and style matching. This external knowledge database contains high-quality speech samples in multiple contexts, covering different emotional expressions, tone changes, and other features. Our experimental results show that our method significantly improves the usability of the speech generation system and the coherence of the generated speech, indicating that our approach has high value in practical applications.

%% file: sections/method.tex
%lwq1223 这里是否考虑给method做个总述？ok
The details of our proposed RAG-based automatic style matching approach are presented in the following sections, including the overall architecture as well as its key modules: the style and timbre decoupled TTS module, the construction and retrieval of the speech style knowledge database, and the PER-LLM-Embedder,an embedding extractor we designed and trained.

\subsection{overall architecture}
\begin{figure}
    \centering
    \includegraphics[width=1.0\linewidth]{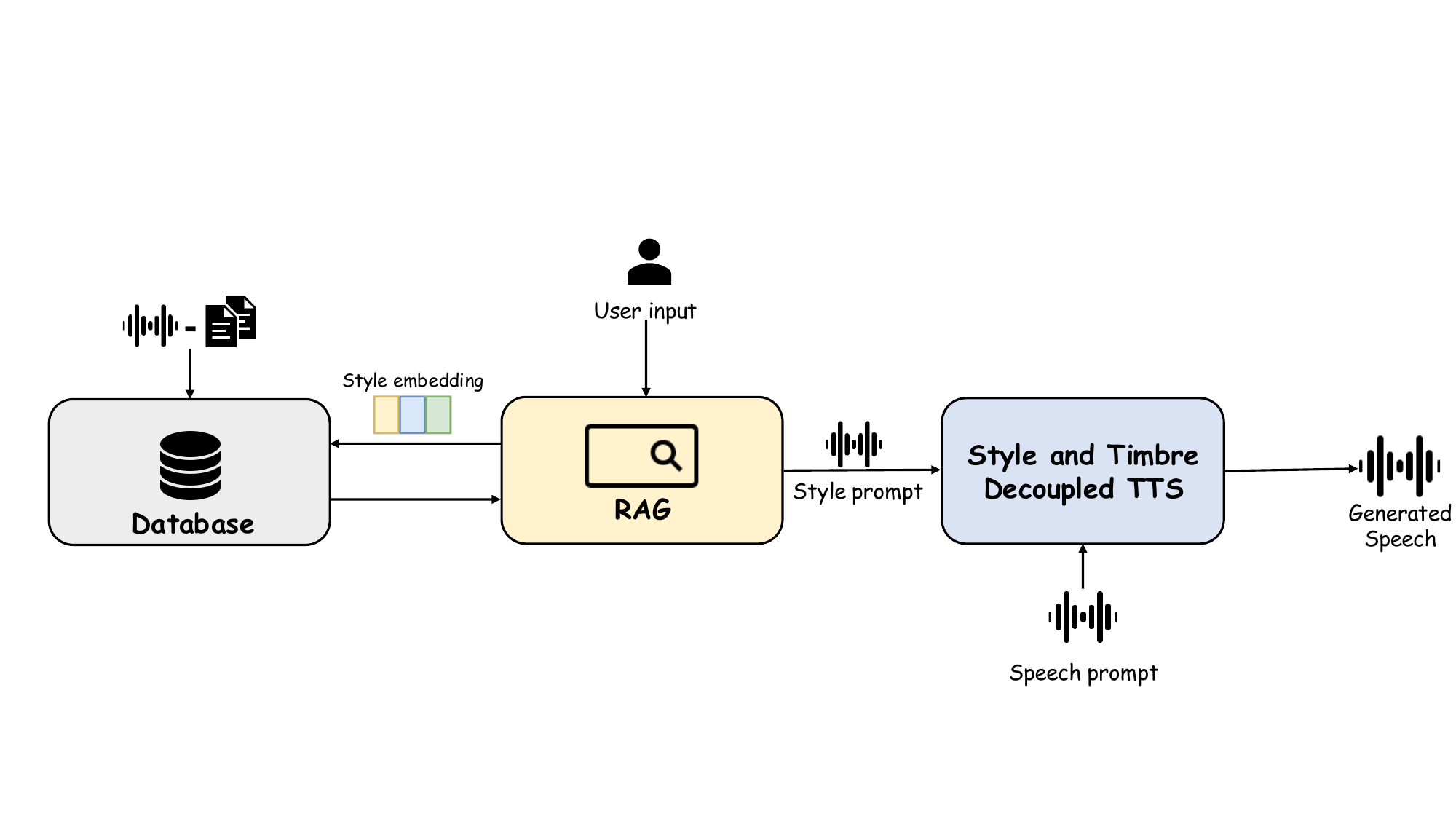}
    \caption{Overall Architecture}
    \label{fig:overall}
\end{figure}
The overall model architecture, as shown in Fig.~\ref{fig:overall} , mainly includes three modules:
\begin{itemize}
    \item Speech Style Knowledge Database: The knowledge database is constructed from a text and speech corpus, consisting of high-quality speech segments containing different emotional expressions, tone changes, and so on. Specifically, it consists of data pairs composed of extracted style embedding and corresponding audio.
    \item RAG Style Selection Module: This module is responsible for analyzing style information based on user query input. It retrieves the most relevant audio data from the retrieval knowledge database and organizes the output as a prompt for the next module.
    \item Style and Timbre Decoupled TTS Module: This module receives style prompts from RAG Style Selection Module as well as user-specified zero-shot timbre prompts. It extracts the specified style and timbre to generate the target speech.
\end{itemize} 

\subsection{Style and Timbre Decoupled TTS}
%lwq1223 这部分是不是提前一些好，放第一部分，先介绍TTS。ok

%lwq1223 需要讲下这个模型是怎么实现解耦的：即用spk emb提供音色信息，用speech token提供风格信息，只在第一部分加入风格emb，而在两个部分都加入了spk emb。“我们分别从timbre prompt和style prompt提取音色和风格信息，作为llm的condition来实现风格和音色的解耦”，具体来说，LLM modeling阶段xxx、flow matching阶段xxx。 ok

\begin{figure}
    \centering
    \includegraphics[width=1.0\linewidth]{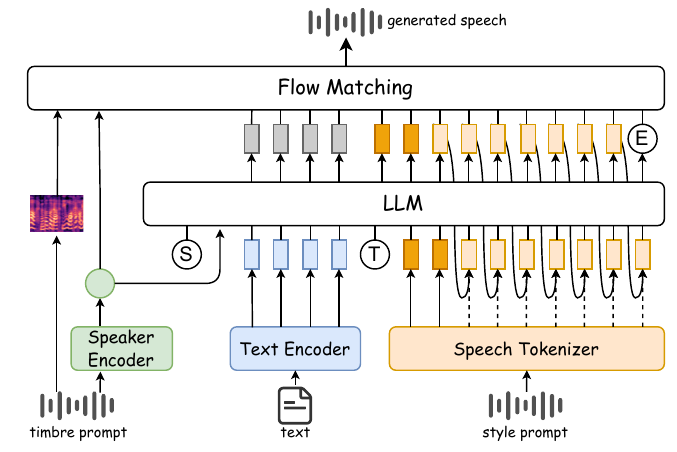}
    \caption{Style and Timbre Decoupled TTS architecture}
    \label{fig:tts}
\end{figure}

%lwq1223 为何speech token存储更多风格信息？和speech token是用whisper的encoder输出有关吗？
In this module, as shown in Fig.~\ref{fig:tts}, we used CosyVoice as the backbone and modified the input of some of its modules. In the frontend, The timbre information is provided by the speaker embedding obtained through the Speaker Encoder, CAM++~\cite{cam++}, while the content and style information are provided by the speech tokens obtained through the speech tokenizer. The speech generation task is modeled in two major parts: LLM modeling and flow matching stage. The style information is injected only in the LLM modeling stage, while the speaker embedding is added in both parts to manage the global vector information.

In the LLM modeling stage, the input sequence is constructed as $[(S), v, \{t\}_{i\in[1:I]}, (T), \{x\}_{k\in[1:K]}, (E)]$, where $(S)$, $(T)$, and $(E)$ represent the start of the sequence, the transition point between text tokens and speech tokens, and the end of the sequence, respectively. $v$ represents the speaker embedding vector extracted from the prompt speech by the speaker encoder, which stores more timbre information; $\{t\}_{i\in[1:I]}$ represents the tokens obtained from the text to be synthesized by a text encoder; $\{x\}_{k\in[1:K]}$ represents the tokens extracted from the speech by the speech tokenizer, which stores more style features. The objective function for training the LLM uses cross-entropy loss:
\begin{equation}
    \mathcal{L}_{LLM}=-\frac{1}{K+1}\sum_{k=1}^{K+1}\text{log}q(x_k)
\end{equation}
where $q(x_k)$ represents the posterior probability of $v_i$ predicted by the LLM and the softmax layer.

The flow matching part models the generation of speech mel-spectrograms conditioned on the aforementioned generated speech token sequences using a flow matching model. The input conditions for the flow matching module include speaker embedding vectors, masked Mel-spectrogram features, and speech token output by the LLM module. Given these input conditions, the flow matching model is trained to match its parameters to the vector field of the speech data distribution~\cite{lipman2023flowmatchinggenerativemodeling}.

\subsection{Speech Style Knowledge Database and User Retrieval}
\label{rag_db}
%lwq1223 method不止是要体现我们是怎么做的，还要体现我们这么做的目的。 1.我们为了xxx（目的）设计了风格语音知识库，为RAG提供基础。2.在构建多模态检索知识库时，预处理、嵌入表示类型和向量数据库类型的重要性（补充下为什么重要），所以我们按照如下方法构建风格语音知识库。3.首先，为了提高xxx准确性、避免长度问题，我们进行了分块，将原始数据中的语音分成x~x秒的块，具体来说，我们的预处理流程如图2所示，xxx。4.接着，我们设计并训练了LLM-embedder。这部分有点混乱，这里的三个LLM是单独微调还是？每个服务一种信息？然后建议是分开说 比如为了保持言语的一致性，我们引入了角色档案信息作为全局信息，该信息是怎么来的（数据流）；为了xx，我们引入xx信息。

In order to enable the support system to flexibly adjust the speech style based on the input text content and provide a more natural and context-appropriate auditory experience, we designed and built a speech style knowledge database, which provides a solid foundation for RAG. This knowledge database stores high-quality speech samples in various contexts, covering features such as different emotional expressions and tone variations.

\begin{figure}
    \centering
    \includegraphics[width=1.0\linewidth]{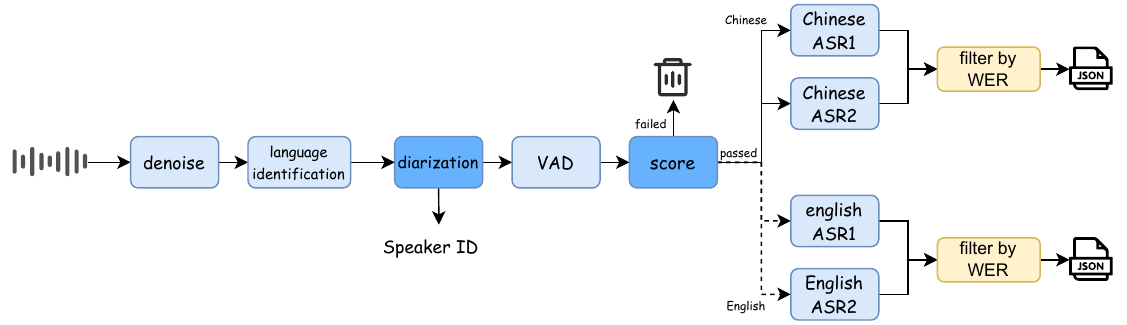}
    \caption{Speech Preprocess Pipeline}
    \label{fig:data_preprocess}
\end{figure}

When constructing the retrieval knowledge base, preprocessing and the type of embedding representation will directly impact the retrieval efficiency and matching accuracy of the system~\cite{10.1145/3178876.3186033}. Good preprocessing can improve data consistency and quality, reducing the impact of noise on retrieval results. Choosing an appropriate embedder model can capture the desired style information for matching, ensuring that the most contextually appropriate speech style samples can be found quickly and accurately. Therefore, we constructed the style speech knowledge base according to the following steps:

Firstly, to improve retrieval accuracy and avoid issues caused by overly long speech segments, we performed chunking on the raw data, dividing each speech sample into short segments of 5 to 10 seconds. This chunking strategy not only helps enhance retrieval precision but also better accommodates text inputs of varying lengths. Specifically, our preprocessing workflow is shown in Fig.~\ref{fig:data_preprocess}, which includes noise reduction, speaker diarization, and other processing. Then, we trim based on Voice Activity Detection (VAD), and segments with scores higher than a preset quality threshold are transcribed into corresponding text data through Automatic Speech Recognition (ASR) and saved into structured data. Next, we chunk the text and audio corpus into smaller blocks. 

\begin{figure*}
    \centering
    \includegraphics[width=1.0\linewidth]{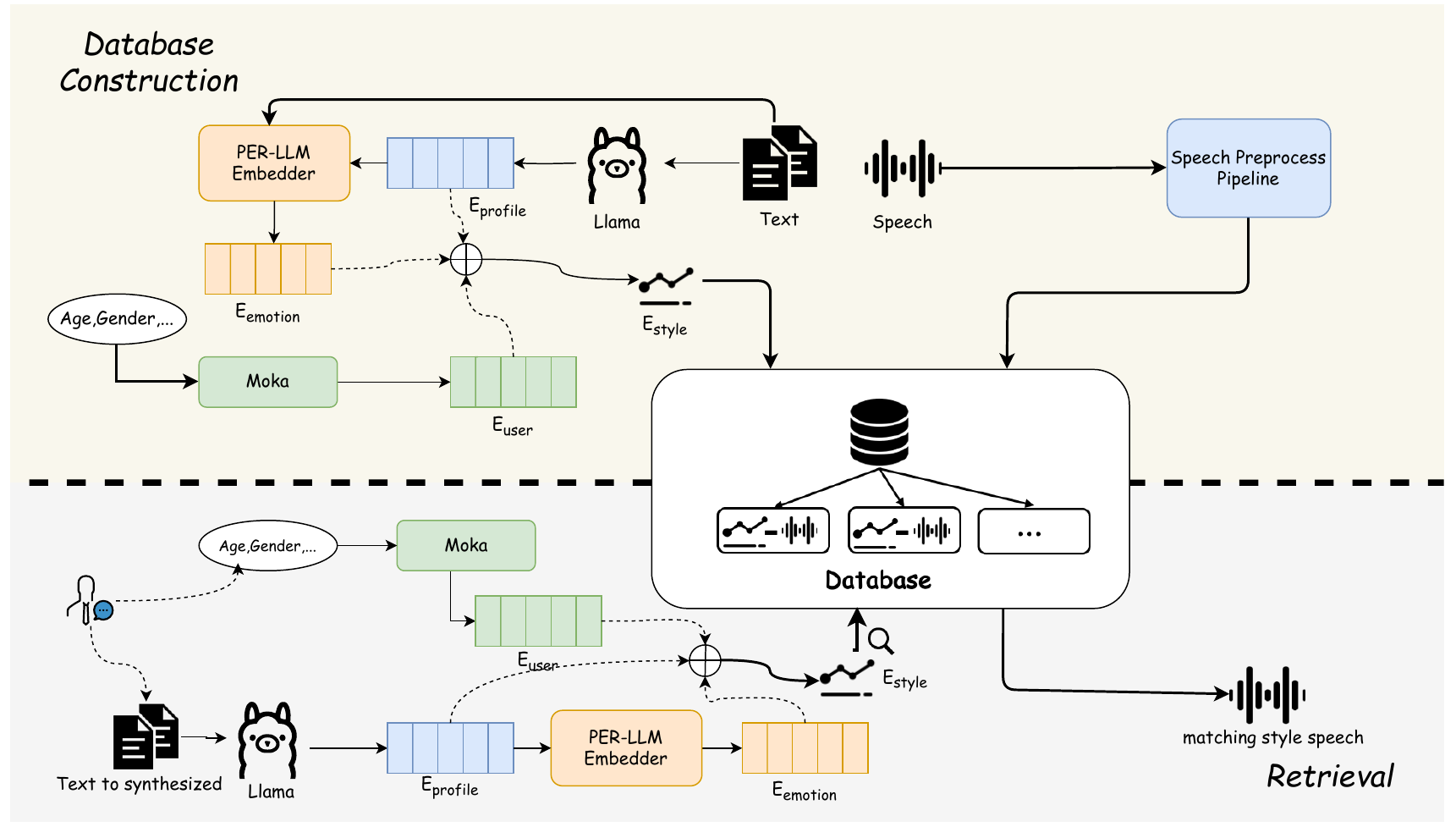}
    \caption{Speech Style Knowledge Database Construction and Retrieval Details}
    \label{fig:db_re}
\end{figure*}

%lwq1223 这里具体用的是lamma3.2还是qwen2.5需要明确下 ok
Next, we designed an embedding extraction module, which consists of the Llama3.2,PER-LLM-Embedder and Moka, which is a industry-proven LLM-embedder. The embedding extraction and user retrieval data flow can be seen in the Fig.~\ref{fig:db_re}. To maintain style consistency, we introduced character profiling as global information, which is extracted by the Llama3.2 from the all of synthesized text. To match the text content with the speech style, we introduced situational emotional information, which is analyzed by the PER-LLM-Embedder from the text and character profiling information. Considering that users may specify their desired style during actual use, we introduced user preference information, which is encoded by Moka based on factors such as age, gender, and region. Finally, the style embedding is extracted for the text content and other preference information. The embedding consists of three parts, as shown in the formula:
\begin{equation}
E_{style}=E_{profile}+E_{emotion}+E_{user}
\end{equation}

Use the combination of these three embeddings for style matching, we can maintain the coherence of speech, change accordingly with the change of speaking context, and provide greater usability of the system.

These speech style embeddings and speech clips are organized into a searchable database, which is indexed using Milvus. The data used in our speech style database includes part of the Microsoft EXPRESSO dataset~\cite{nguyen2023expressobenchmarkanalysisdiscrete}, which has expressive english speech and other high-quality expressive Chinese speech. And ultimately, it consists of high-quality speech clips, including 30 speakers and more than 2000 speech segments.

% \subsection{RAG Style Retrieval}
%lwq1223 这段有点啰嗦了，用户提供prompt就不检索 没提供就检索？ 一句话就能概括了。这段建议重申下RAG的作用。
In the retrieval stage, we first judge the user's query to determine whether retrieval is necessary. Then we proceed to further retrieval enhancement processing.
The way we carry out retrieval enhancement processing is to select the top k most relevant to the query from the pre-constructed knowledge database, based on the similarity of embeddings. The specific processing method is: firstly, we rewrite the query based on the user's input information; , secondly,same as database construction,use the Llama, PER-LLM-Embedder, Moka to get embedding. Lastly, we use Max Inner Product Search (MIPS)~\cite{auvolat2015clusteringefficientapproximatemaximum} to calculate the similarity and obtain the final top-K prompts.

\subsection{PER-LLM-Embedder Training}
Since we want to obtain embeddings that represent situational emotions, and currently LLM perform exceptionally well in understanding the semantics and emotions of text, we fine-tune LLaMA3.2 on a specific dataset. We utilized the IEMOCAP~\cite{busso2008iemocap} dataset, which consists of daily conversational scenarios and scripted emotional performances from ten actors, as well as the Multimodal Multiscene Multilabel Emotion Dialogue (M3ED)~\cite{zhao-etal-2022-m3ed} dataset in Chinese, comprising 990 binary emotional dialogue video clips extracted from 56 distinct television dramas (500 episodes in total). These dataset have some dialogue text with speaker-id and the emotion labels. During the fine-tuning stage, we input dialogue text and character profiles into the model with the goal of predicting emotion labels. Specifically, the English data for the character profiles is generated by LLaMA 3.2, while the Chinese data is generated by Qwen2.5.
Regarding the implementation details, we employed a learning rate of $5e-4$, a dropout rate of $0.2$, and set the number of epochs to $3$. Additionally, the local context window size $(w)$ was set to 5.

%% file: sections/exp.tex
\subsection{Experimental Setup}
%lwq1224 IEMOCAP是用来训练哪个模块需要说明一下

\textbf{Evaluation metrics.} 
Our evaluation metrics are divided into objective and subjective evaluations.
For objective metrics, we considered five evaluation indicators: Speaker Similarity (SIM), Word Error Rate (WER), 
 Virtual Speech Quality Objective Listener(VISQOL), KL divergence(KL), and Inception score(IS). SIM evaluates the similarity between the generated audio and the reference audio. WER calculates the word error rate between the transcribed text of the generated audio and the reference text. VISQOL is an objective, full-reference metric for perceived audio quality. KL quantifies the difference between generated speech and ground truth; IS evaluates the diversity and realism of speech generated by a model.
 
For subjective metrics, we conducted two Mean Opinion Score (MOS) subjective tests: Style Matching MOS(SM-MOS) and Style Coherence MOS(SC-MOS). SM-MOS evaluates the style matching degree between the text and the synthesized speech, while SC-MOS is used to compare whether the overall style of the synthesized speech is harmonious and whether the emotional fluctuations are appropriate.
Additionally, we set up an AB test to evaluate the effectiveness of our method compared to manual selection.

\subsection{Style and Timbre Decouple Experiment}
\begin{figure}
    \centering
    \includegraphics[width=1.0\linewidth]{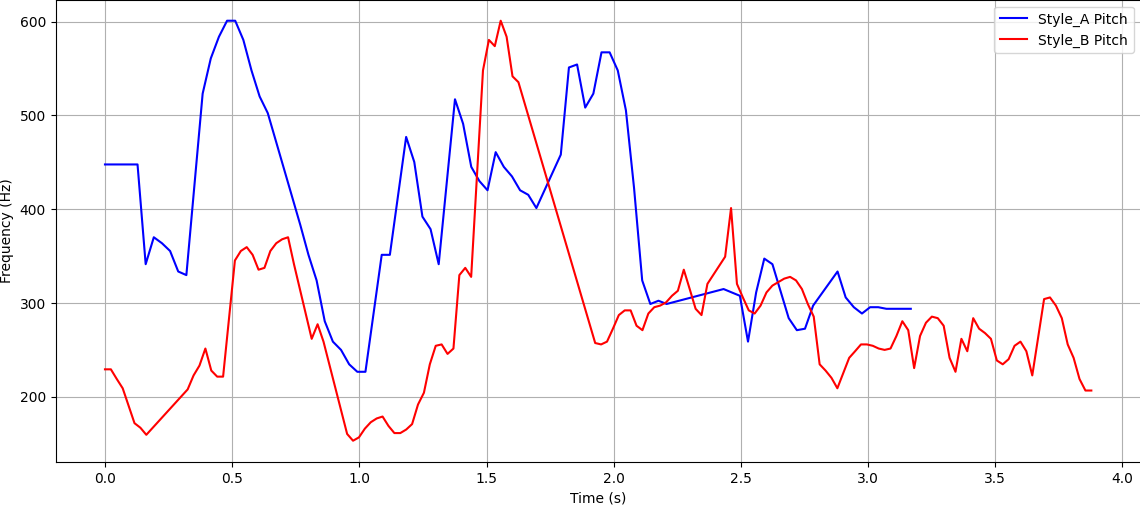}
    \caption{Different Style Prompt Speech Synthesis Result Visualization}
    \label{fig:mel}
\end{figure}

To demonstrate that our model can transform speech styles without affecting its ability to generate accurate timbre and word correctness, we employ 1,00 samples from the Common Voice dataset and the DiDiSpeech-2 dataset as the test set. For Cosyvoice, we generate four sentences for each sample; for our method, we generate four different style results for each sample (visualization of the different style speech results can be seen in Fig.~\ref{fig:mel}).
We use WER, SIM, VISQOL, KL and IS metrics for evaluation. 

\begin{table}[htbp]
\caption{Style and Timbre Decouple Experiment Result}
\begin{center}
\begin{tabular}{|c|c|c|c|c|c|}
\hline
\textbf{Model}&\multicolumn{5}{|c|}{\textbf{metrics}} \\
\cline{2-6} 
\textbf{name} & \textbf{\textit{SIM}$\uparrow$}& \textbf{\textit{WER(\%)}$\downarrow$} & \textbf{\textit{VISQOL}$\uparrow$} & \textbf{\textit{KL}$\downarrow$}  & \textbf{\textit{IS}$\uparrow$}\\
\hline
 CosyVoice~\cite{du2024cosyvoice}& 0.753  & 2.312 &  4.083 & 0.165 & 1.007  \\
Ours & 0.750 &  3.600 & 4.075 & 0.160 & 1.325 \\
\hline
\end{tabular}
\label{tab1e:decouple}
\end{center}
\end{table}

As illustrated in TABLE \ref{tab1e:decouple} ,the results indicate that our method can achieve style control while maintaining timbre and speech quality. Furthermore, the IS metric score demonstrates that our method can generate results with more varied styles and possesses more flexible generation capabilities.

\subsection{RAG Impact Evaluation}
\label{main_exp}
We first conducted a subjective evaluation of the backbone module of the TTS system, CosyVoice~\cite{du2024cosyvoice}, another state-of-the-art model in the TTS field, MaskGCT~\cite{wang2024maskgct}, and our method, using the SM-MOS and SC-MOS metrics and inviting 15 volunteers to participate in the evaluation. The test set was divided based on language, with data for each language 12 audio samples covering both story narration and dialogue scenarios. For each sample, CosyVoice used the generated speech from the previous sentence as the speech prompt for the next sentence.

\begin{table}[htbp]
\caption{rag impact subjective eval}
\begin{center}
\begin{tabular}{|c|c|c|c|}
\hline
\textbf{Model}& \multicolumn{1}{|c|}{\textbf{\textit{Language}}} & \multicolumn{2}{|c|}{\textbf{Metrics}} \\
\cline{3-4} 
\textbf{name} &  & \textbf{\textit{SM-MOS}$\uparrow$} & \textbf{\textit{SC-MOS}$\uparrow$} \\
\hline
CosyVoice~\cite{du2024cosyvoice} &  & 3.35±0.13 & 3.48±0.13 \\
MaskGCT~\cite{wang2024maskgct} & English  & 3.85±0.13 & 3.81±0.12 \\
AutoStyle-TTS(Ours) &  & \textbf{3.85±0.13} & \textbf{3.81±0.12} \\
\hline
CosyVoice~\cite{du2024cosyvoice} &  & 3.38±0.14 & 3.51±0.14 \\
MaskGCT~\cite{wang2024maskgct} & Chinese & 3.85±0.13 & 3.81±0.12 \\
AutoStyle-TTS(Ours) &  & \textbf{3.90±0.12} & \textbf{3.83±0.13} \\
\hline
\end{tabular}
\label{tab1}
\end{center}
\end{table}

According to the experimental results, our method significantly outperformed Cosyvoice on both key metrics, indicating that our technology has made substantial progress in text-style matching and overall speech coherence.

\begin{figure}
    \centering
    \includegraphics[width=1.0\linewidth]{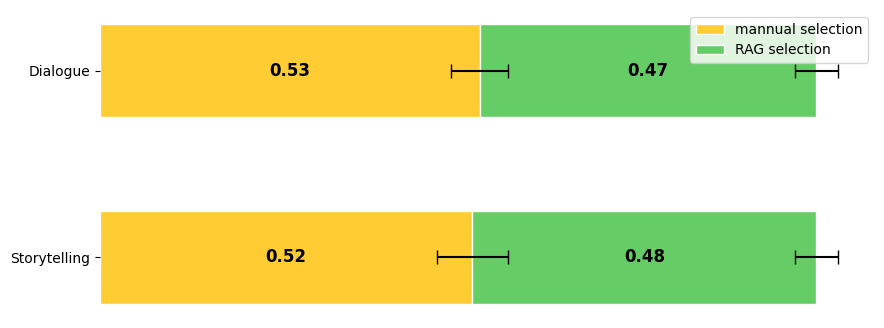}
    \caption{AB Test Result}
    \label{fig:abtest}
\end{figure}

Secondly, the evaluation also included an AB preference test, where participants were asked to compare the speech styles selected by our method with those chosen manually, according to a certain requirement. Specifically, we performed speech synthesis for 10 text passages(include storytelling and dialogue) using both manually selected and model-automatically selected speech styles.
Thirty participants were asked to choose their preferred samples.
According to the results of the AB test in Fig.~\ref{fig:abtest}, our method is comparable to the manually selected prompt method in terms of user preference. The experimental results demonstrate that our proposed automated style matching mechanism can effectively replace the time-consuming manual selection process while ensuring that the synthesized speech's style is highly consistent with the content, thereby providing a natural and coherent auditory experience.

\subsection{Ablation Experiment}

To validate the effectiveness of the character profiling and situational emotion in the style embeddings extracted by Llama and our PER-LLM-Embedder, we conducted a series of ablation experiments. The metrics, number of test participants, and number of test samples per group are the same as~\ref{main_exp}. The profile-only and emotion-only represent the cases where only character profiling or situational emotion was used for similarity calculation to match the style prompt. The K represents the number of retrieved style prompts, and we concat the style prompts together and enter them into the subsequent TTS module.

\begin{table}[htbp]
\caption{Ablation Experiments for Embedding and Top-k}
\begin{center}
\begin{tabular}{|c|c|c|c|}
\hline
\textbf{Ablation}& \textbf{Retrieval} & \multicolumn{2}{|c|}{\textbf{metrics}} \\
\cline{3-4} 
\textbf{Module} & \textbf{Method} & \textbf{\textit{SM-MOS}$\uparrow$}& \textbf{\textit{SC-MOS}$\uparrow$} \\
\hline
 & only-profile &   3.40±0.12   & 2.91±0.13   \\
Embedding& only-emotion  &   3.20±0.12   & 3.60±0.14   \\
 & profile+emotion(ours) &     \textbf{3.85±0.13}   & \textbf{3.81±0.12} \\
\hline
 & K = 1 &    3.65±0.14   & 3.61±0.13 \\
Top-K& K = 3  &    \textbf{3.85±0.12}   & \textbf{3.81±0.13} \\
 & K= 5 &    3.75±0.13   & 3.61±0.14 \\
\hline
\end{tabular}
\label{tab1e:ablation}
\end{center}
\end{table}

The results indicate that the model achieves the best performance when matching with both character profiles and situational emotions. 
The method that uses only character profile embeddings for matching shows a decline in the SM-MOS metric. This is because this method loses the situational emotional information of individual sentences, making it difficult to match the emotions of the text with those of the synthesized speech, leading to a decrease in the style matching score. Also, This approach shows a decline in the SC-MOS metric. This is because the loss of emotional matching leads to disjointed transitions in the overall emotional fluctuations.
For the approach that uses only situational emotions, the main decline is observed in the SC-MOS metric. The absence of comprehensive character profile information for control results in some loss of overall coherence. However, the decline is not significant because the situational emotional information is extracted with the aid of character profile information, which carries some global information, helping to maintain the coherence of the speech.

To evaluate the impact of the number of speech prompts on the RAG method for speech style, we conducted ablation experiments, as shown in the TABLE~\ref{tab1e:ablation}. We found that as the number of prompts increased, the speaker similarity also gradually increased. This is because the TTS system attempts to clone the speaking style, and longer style prompts provide more appropriate style information. However, the results also indicate that performance peaks at a prompt length of 3 and diminishes with further increases in prompt length. We believe that longer, inconsistent style prompts from different sources may hinder the TTS system's ability to generate coherent speech.